# Wafer-scale fabrication of two-dimensional β-In$_2$Se$_3$ photodetectors

*Marcel S. Claro [a,b]\*, Justyna Grzonka [a], Nicoleta Nicoara [a,b], Paulo J. Ferreira [a,c,d], Sascha Sadewasser [a,b]*

[a] International Iberian Nanotechnology Laboratory, Av. Mestre José Veiga, 4715-330, Braga, Portugal

[b] QuantaLab, 4715-330 Braga, Portugal

[c] Mechanical Engineering Department and IDMEC, Instituto Superior Técnico, University of Lisbon, Av. Rovisco Pais, 1049-001 Lisboa, Portugal

[d] Materials Science and Engineering Program, The University of Texas at Austin, Austin, Texas 78712, USA.

*The epitaxial growth of two-dimensional (2D) β-In$_2$Se$_3$ material was obtained over 2-inches c-sapphire wafers using molecular beam epitaxy (MBE). Excellent quality of thick (90 nm) and very thin films, down to two quintuple layers (2 nm), was confirmed by X-ray diffraction (XRD), Raman spectroscopy, and aberration-corrected scanning transmission electron microscopy (ac-STEM). Wafer-scale fabrication of photodetectors based on five quintuple layers was produced using photolithography and other standard semiconductor processing methods. The photodetectors exhibit responsivity of 3 mA/W, peak specific detectivity (D\*) of 10$^9$ Jones, external quantum efficiency (EQE) of 0.67% at 550 nm, and response-time of ~7 ms, which is faster than any result previously reported for β-In$_2$Se$_3$ photodetectors. From the photocurrent measurements, an optical bandgap of 1.38 eV was observed. These results on wafer-scale deposition of 2D In$_2$Se$_3$, as well as its fabrication into optoelectronic devices provide the missing link that will enable the commercialization of 2D materials.*

The large family of two-dimensional (2D) materials is considered the next generation of electronic materials. They have been investigated extensively in the past years, motivated by their novel electronic and optical properties[1], which can be tuned by simply changing the number of layers in a given material[2]. Nevertheless, the research of 2D materials is still in its early stages. The field started with graphene several years ago[3], but more materials have been added continuously to the list. Among these, transition-metal di-chalcogenides (TMDC)[4], which have impressive optical properties in the visible range, are currently one of the most studied 2D materials. Devices made of these materials have shown remarkable performance as photodetectors[5,6], non-volatile random-access memory[7] or photovoltaics[8], even when using simple manual exfoliation methods.

More recently, 2D indium-selenides (In$_x$Se$_y$) have gained attention[9], due to their bandgap in the visible spectral region, comparable to TMDCs, but also due to several novel properties: InSe exhibits one of the largest mobilities of 2D semiconductor materials[10], and β-In$_2$Se$_3$ shows good mobility[11], excellent photoresponsivity[12], and exotic ferroelectricity[13]. Transistors with high mobility and on/off ratio have already been realized.[14] Therefore, 2D In$_2$Se$_3$ materials have the potential to address several limitations of the current silicon (Si) and III-V technologies, such as improved mobility and overall performance of transistors for electronics, as well as integrated photodetectors and light emitters in the same material system (same die), all possible on virtually any substrate, including transparent and flexible substrates.[15,16] Yet, nearly all reported devices from 2D materials up to now rely on fabrication methods based on exfoliation and transfer of layers onto other substrates or other 2D materials. While this device fabrication process allows unprecedented flexibility in the combination of materials and therefore has nearly unlimited device design possibilities[1], it leads typically to individual or few devices and is a slow and tedious process.

On the other hand, due to their layered nature, where van der Waals forces exist between the layers, 2D materials can be grown epitaxially on substrates without strong chemical bonds to the surface, a growth method called *van der Waals epitaxy*[17]. The substrate does influence the growth[18], but the growth can be performed on virtually any substrate without the formation of any strain in the layer. However, so far, only a few large-scale production methods combining van der Waals epitaxy and standard semiconductor processing have been reported [19–21].

In the case of indium selenide, which occurs in several phases and stoichiometries (InSe, α-$In_2Se_3$, β-$In_2Se_3$, $In_3Se_4$ etc.[9]), the synthesis of single-phase materials is challenging. Yet, among all the variety of 2D crystal growth techniques, molecular beam epitaxy (MBE) is the process that is capable of producing materials with exceptional purity and controllability of their physical properties[22].

In this work, we demonstrate that MBE growth processes can be fine-tuned to obtain phase-control and produce few layers of 2D materials at wafer scale. Few layers of β-$In_2Se_3$ were applied in the fabrication of hundreds of photodetectors on a full wafer using standard semiconductor processing. They exhibited good detectivity and the fastest response times so far observed for this material. These results establish the foundation for the future commercialization of 2D materials.[23]

The growth of $In_2Se_3$ was realized in a MBE by independently evaporating indium from a Knudsen cell and selenium from a valved cracker cell, where the Se flux is controlled by the valve aperture ranging from 0 to 8 mm. The growth is performed on pre-annealed 2-inch epi-ready single-side polished c-sapphire (0001) substrates.

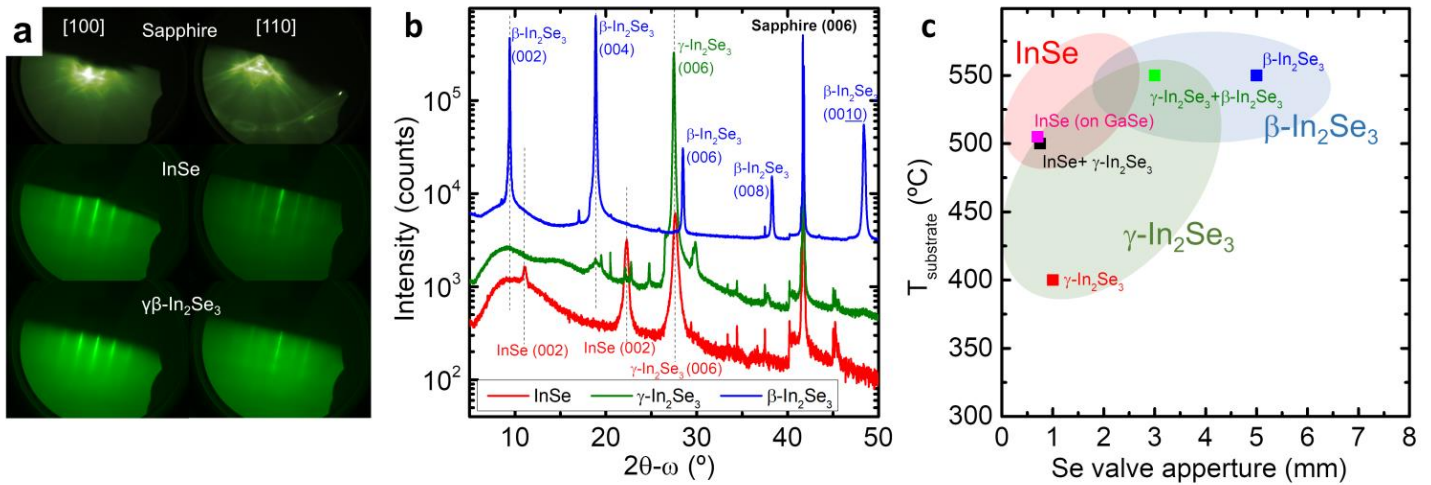

**Figure 1.** Molecular beam epitaxy growth of Indium Selenide. **a** In-situ reflection high-energy electron diffraction (RHEED) pattern of the sapphire substrate and the different indium-selenide phases obtained. **b** X-ray diffraction (XRD) pattern of the different phases obtained on sapphire substrates. **c** Proposed phase diagram for the MBE growth of $In_xSe_y$.

To identify the growth conditions for single-phase β-$In_2Se_3$, the substrate temperature and the In/Se ratio were varied. The Indium source was maintained at 770 ºC (corresponding to an In growth rate of ~1.5 nm/min), while the Se valve aperture was varied between 0.5 and 5 mm. The substrate temperature was adjusted between 300 and 600 ºC. It was found that the best growth temperatures concerning surface roughness and single-phase are between 500 and 550 ºC. Lower temperatures produced rough surface and polycrystalline films, and at higher temperatures $In_xSe_y$ is desorbed from the surface. In this temperature window, three phases of $In_xSe_y$ were obtained by varying the valve aperture. InSe was obtained at 0.75 mm or less, γ-$In_2Se_3$ from 1 to 3 mm and β-$In_2Se_3$ from 2 to 5 mm. For some valve apertures a phase mixture was obtained while for specific conditions single-phase material were obtained. The growth was monitored in-situ by reflection high-energy electron diffraction (RHEED), which distinguishes between the initial sapphire substrate and the different phases obtained under different growth conditions (Fig. 1a). InSe exhibits a very distinct surface diffraction pattern from the γ- or β-$In_2Se_3$ phases, while γ- and β-$In_2Se_3$ RHEED patterns are indistinguishable during the growth. For growth processes showing only one of the distinct patterns, ex-situ techniques are required to properly determine the phase obtained. Considering that β-In$_2$Se$_3$ has a broad and weak Raman signal, we use X-ray diffraction (XRD) to distinguish between the phases after growth (Fig. 1b).

In Fig. 1c, we show a phase diagram for the MBE growth of the various In$_x$Se$_y$ phases, based on our XRD data and RHEED observations. We should emphasize that the substrate is fundamental for the determination of the phase that grows, as illustrated by the growth of the single-phase InSe on *c*-sapphire with a GaSe buffer layer, versus the formation of mixed InSe and γ-In$_2$Se$_3$ phases on *c*-sapphire, in agreement with a previous report[24]. In short, to obtain single-phase InSe, In-rich growth conditions are required, while strongly Se-rich conditions lead to single-phase β-In$_2$Se$_3$. For intermediate conditions or a large substrate lattice mismatch, the 3D γ-In$_2$Se$_3$ phase is obtained.

From the phase diagram we conclude that the optimized growth conditions for single-phase β-In$_2$Se$_3$ is for a substrate temperature of 550 ºC and 5 mm Se valve aperture. Using X-ray reflection (XRR) measurements, a layer thickness of 90 nm was determined for 60 min of growth, corresponding to a growth rate of 1.5 nm/min. Atomic force microscopy (AFM) measurements show height variations of up to 10 layers and an RMS roughness of ~1.34 nm (Fig. 2a-b). Pyramidal shapes of In$_2$Se$_3$ terraces pointing in two opposite directions indicate the presence of twinning, and their spiral shape at the top shows screw-dislocation-driven growth, similar to reports for other layered materials[25,26]. A similar amount of both twinned orientations is observed by an XRD phi-scan of the ($11\bar{2}0$) peak (Fig. S1 [Suppl. Information]).

Raman spectroscopy (Fig. 2c) shows clear peaks at 110.0 cm$^{-1}$ ($A_1$ mode), 175.1 cm$^{-1}$ ($E_g$ mode), and 206.9 cm$^{-1}$ ($A'_1$ mode), indicating β-phase In$_2$Se$_3$.[27] This spectrum, mainly the position of the first $A_1$ peak, is notably different from the InSe ($A_1$=115 cm$^{-1}$) and γ-In$_2$Se$_3$ ($A_1$=152 cm$^{-1}$) signatures[28], however, it is close to that of α-In$_2$Se$_3$[24], as in the XRD pattern[29]. Nevertheless, the presence of the α-phase was not observed by any other technique, including aberration-corrected scanning transmission electron microscopy (ac-STEM) images (see below). The Raman signal of the β-phase is weaker compared to the other phases; however, attempts to improve the signal by increasing the laser power usually resulted in sample modification as observed by peaks corresponding to amorphous Se[30] and to γ-In$_2$Se$_3$, induced by local heating (see also the discussion of the Raman measurements of a 5 QL thick sample below). The $A_1$ peak of β-In$_2$Se$_3$ is notably broader (FWHM ~20 cm$^{-1}$) than usual peaks of other 2D materials[31]. Nevertheless, such broad $A_1$ peaks have been observed previously[27]; therefore, they reflect an inherent material property and are not related to the quality of our specific material.

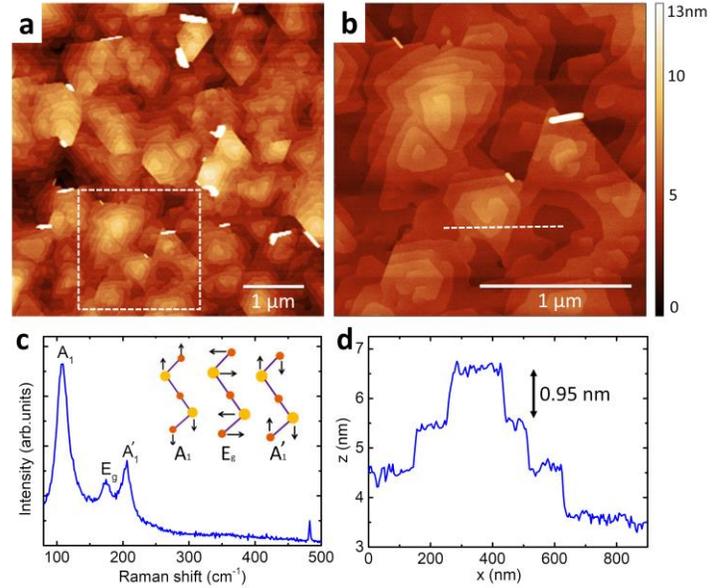

**Figure 2.** Characterization of 90 nm thick single-phase β-In$_2$Se$_3$ grown on *c*-sapphire. **a** Atomic force microscopy (AFM) of the surface in a 5 × 5 μm$^2$ area. The white dashed-line square highlights the area shown in b. **b** Higher magnification AFM image with 2 × 2 μm$^2$. The dashed white line indicates the position of the height profile shown in **d**. **c** Raman spectrum taken at 8 mW power at the objective (2 s acquisition time). The inset illustrates the identified vibrational modes. **d** Height profile along the dashed white line indicated in **b**, showing individual steps of ~0.95 nm height, corresponding to single quintuple layers (QL).

To assess the crystalline quality of the layers, high-angle annular dark-field (HAADF) ac-STEM was performed. At low magnifications, a wide-range (1.2 μm length) cross-sectional view of our MBE-grown 90-nm thick β-In$_2$Se$_3$ film on *c*-plane sapphire (Fig. 3a) shows no visible defects, as twining or grain boundaries. The HAADF ac-STEM images clearly distinguish the In$_2$Se$_3$ film (bright contrast) from the sapphire substrate (darker contrast), as this imaging mode is strongly dependent on the atomic number (~$Z^2$). The selected area electron diffraction (SAED) pattern (Fig. 3b), obtained from the interface between the substrate and the film, shows that β-In$_2$Se$_3$ is single-crystalline and that it grows epitaxially on *c*-plane sapphire with the following crystallographic relationships: $In_2Se_3$ ($11\bar{2}0$) ∥ sapphire ($11\bar{2}0$) and $In_2Se_3$ (0001) ∥ sapphire (0001). The structural ordering of the film is seen at higher magnifications (Fig. 3c and d), which shows the characteristic QL structure corresponding to the β-$In_2Se_3$ phase[32]. In particular, the interface region between β-$In_2Se_3$ and sapphire shows that the film is very well ordered during the initial stages of growth. The layered structure is clearly visible as shown by the van der Waals gap separating each set of five atomic layers Se-In-Se-In-Se (Fig. 3d). Along the selected [$11\bar{2}0$] zone axis, pure In and Se atomic columns prevail, which are thus easily distinguishable due to the higher atomic number of In compared to Se. HAADF ac-STEM atomic scale images of a single QL clearly shows the atomic stacking of In and Se atomic columns (Fig. 3e), which match well HAADF ac-STEM image simulations using a multislice algorithm (Fig. 3f).

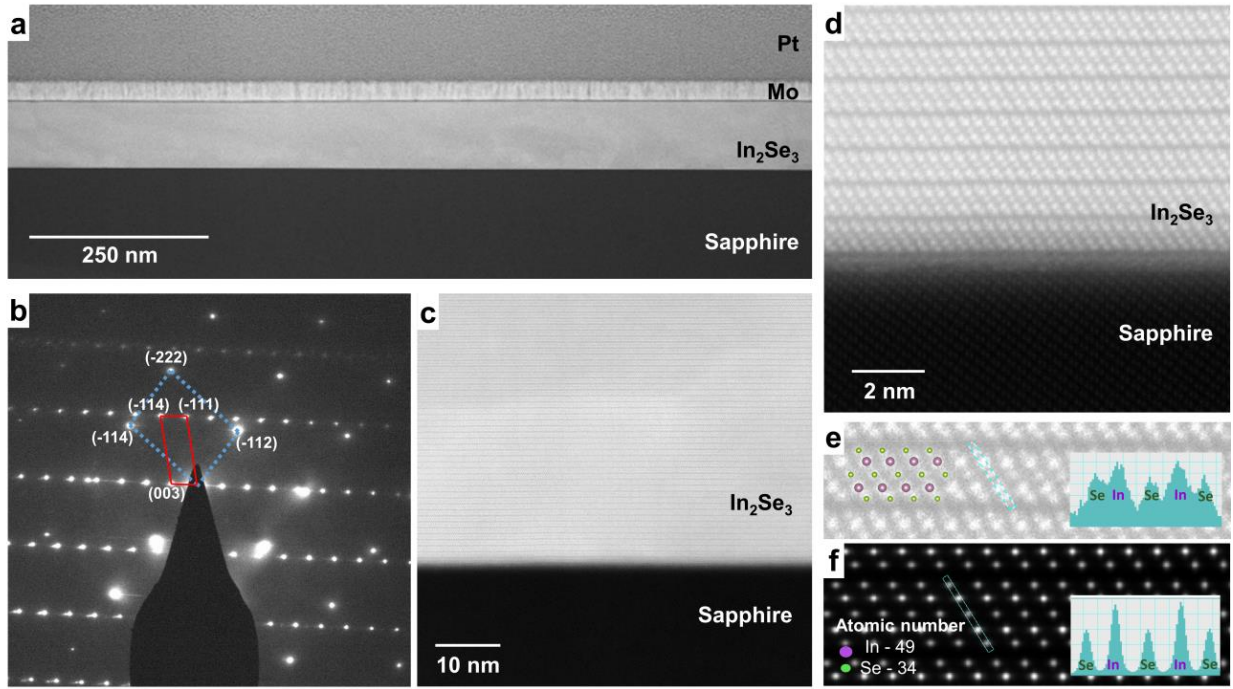

**Figure 3.** HAADF ac-STEM analysis of 90 nm thick β-$In_2Se_3$ grown on *c*-plane sapphire. **a** Low magnification HAADF ac-STEM image of the $In_2Se_3$ layer. **b** SAED from the $In_2Se_3$/sapphire interface region along the [$11\bar{2}0$] zone axis. The dashed (blue) and solid (red) rectangles are showing the projected unit cells of sapphire and $In_2Se_3$ respectively. **c** HAADF ac-STEM image showing the well-ordered layer structure of $In_2Se_3$ grown on saphire. **d** Higher magnification HAADF ac-STEM image showing the smooth $In_2Se_3$/sapphire interface and regular $In_2Se_3$ quintuple layers. **e** HAADF ac-STEM image at the atomic scale showing a single quintuple layer with an overlaid structural model and intensity profile. **f** Multislice HAADF ac-STEM image simulation of quintuple $In_2Se_3$ layers and intensity profile showing a good match with the experimental data obtained in **e**. The strongly scattering (bright) atoms correspond to In and the ones with lower brightness to Se.

Following the same growth conditions of the thick films, we produced thinner films of five QLs, using a growth time of 90 s. AFM images indicate a very flat film with only three terrace levels around the average thickness of 5 QLs (Fig. S2a [Suppl. Information]), confirming that a continuous and uniform film of 5±1 QLs across the 2-inch wafer was obtained, with reproducibility over many growth processes. The β-$In_2Se_3$ structure is confirmed by XRD measurements (Fig. S2c [Suppl. Information]) and Raman spectroscopy (Fig. S2d [Suppl. Information]).

To demonstrate the capability to grow even thinner β-$In_2Se_3$ films, the growth rate was reduced to 1 QL in ~390 s, by reducing the indium effusion cell temperature to 670 ºC. This growth rate was determined in-situ by measuring the time necessary for the RHEED pattern to change from the one associated with *c*-sapphire to that for $In_2Se_3$ (see Fig. 1a). However, due to the slow growth, it was not possible to observe further RHEED oscillations at these long times. Here, we show results for a sample with only 2 QL grown by this process with a duration of 13 min. The characterization of samples with few QLs is very challenging, since such thin layers show very weak Raman and weak and broad XRD signals, which are difficult to separate from the background and noise in our experimental setups. AFM images taken directly on the In$_2$Se$_3$ surface do not show clear atomic terraces, but a rather smooth surface. We attribute this difficulty in AFM imaging to the weak van der Walls binding forces at the interface between the substrate and In$_2$Se$_3$ layers. Nonetheless, we succeeded to observe clear atomic terraces after covering the surface with 10 nm of sputtered amorphous Al$_x$O$_y$, which conforms perfectly to the surface (Fig. 4a). As in the case of the 5 QL thick sample, three levels of terraces are observed, a complete layer (likely the first QL), an almost complete layer and a third layer starting to form. The typical triangular terraces with ~0.95 nm height can be identified (see Fig. S3 [Suppl. Information]), supporting the growth of hexagonal β-In$_2$Se$_3$. The RMS roughness is 0.42 nm, lower than the QL thickness.

To additionally confirm the growth of beta-phase In$_2$Se$_3$, grazing incidence in-plane XRD (GIIXRD) was performed (Fig. 4b); this technique was already successfully applied for the identification of few-layer WSe$_2$, when out-of-plane XRD fails due to the lack of periodicity in the *c*-axis (00l) of samples with very few layers.[35] Here, we observe a small peak at 46.09 º for the 2 QL sample, which we attribute to the diffraction of the (11$\bar{2}$0) plane of β-In$_2$Se$_3$, in agreement with the observation of the same peak for the 90 nm thick β-In$_2$Se$_3$ sample.

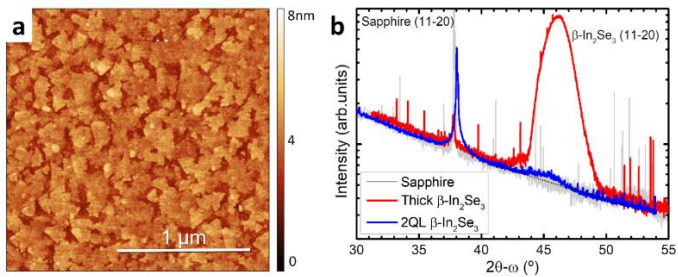

**Figure 4.** Topography and structure of a 2 QL β-In$_2$Se$_3$ sample. **a** Atomic force microscopy (AFM) of the surface after conformal coating with 10 nm Al$_x$O$_y$ by sputter deposition. 2 × 2 µm$^2$. **b** Grazing incidence in-plane XRD (GIIXRD) of the (11$\bar{2}$0) peaks of *c*-sapphire (light grey), 90 nm β-In$_2$Se$_3$ (red), and 2 QLs β-In$_2$Se$_3$ (blue).

Finally, we demonstrate the capability to process the 5-QL β-In$_2$Se$_3$ into devices. Attempts to directly process the β-In$_2$Se$_3$ films by lithography processes led to the formation of In$_x$O$_y$ (see Fig S4a [Suppl. Information]); therefore, the films were protected with 30 nm of sputtered intrinsic ZnO (i-ZnO). The integrity of the β-In$_2$Se$_3$ upon i-ZnO sputtering was confirmed by XRD (Fig. S4b [Suppl. Information]). The i-ZnO is a semi-insulating oxide with sufficient resistance (10$^4$ Ohm/sq sheet resistance for a 50 nm thick film) to work as passivation layer (but not as a gate). More importantly, i-ZnO is easily etched with dilute HCl, while β-In$_2$Se$_3$ is rather resilient to this HCl etch; therefore, HCl provides a very effective selective etching. Other insulating oxides (e.g. Al$_2$O$_3$ or SiO$_2$) would require reactive ion etching or harsher chemical etchants, which would compromise the quality of the β-In$_2$Se$_3$. Furthermore, i-ZnO is also transparent to visible light down to 375 nm wavelength.

Thus, for device fabrication two steps of direct photolithography were utilized. In the first step the i-ZnO is removed for the deposition of the metallic contacts using a solution of aqueous HCl (0.5%) for 15 s. Following, without removing the photoresist, the Cr/Au contacts (thickness of 6 and 30 nm, respectively) were deposited by sputtering immediately after opening the via to prevent oxidation. The unwanted metal film is removed by a lift-off process in acetone. In the second step that defines the channel, the i-ZnO is removed by the same solution, followed by removal of the β-In$_2$Se$_3$ by a more diluted solution of 0.3% HCl with 30% H$_2$O$_2$ added in the proportion of 4:50. H$_2$O$_2$ oxidizes the β-In$_2$Se$_3$, such that it can be removed by the HCl solution, resembling the principle of standard III-V etch solutions. This second etching of the β-In$_2$Se$_3$ occurs in 10 to 15 seconds, which can be easily monitored optically since the orange/yellow color of β-In$_2$Se$_3$ is observable until the layers are completely removed.

This processing method (illustrated in Fig. S5 [Suppl. Information]) is applied to the whole 2-inch wafer with several geometries for different devices, leading to a total of more than 250 devices (Fig. 5a). Here, we present the results of 2-terminal devices (inset of Fig. 5a) with a channel length of 100 µm and two different channel widths of 40 µm and 20 µm. The quality of the processing is validated by ac-STEM examination. Cross-section HAADF ac-STEM images of the channel region with i-ZnO (Fig. 5b) show a well-defined interface between β-In$_2$Se$_3$ and ZnO, reconfirming the integrity of the β-In$_2$Se$_3$ after the sputtering process of i-ZnO. However, in the contact region, the top layers of the β-In$_2$Se$_3$ appear to be affected by the etching (Fig. 5c), suggesting that further process optimization is required to enable devices with fewer QLs.

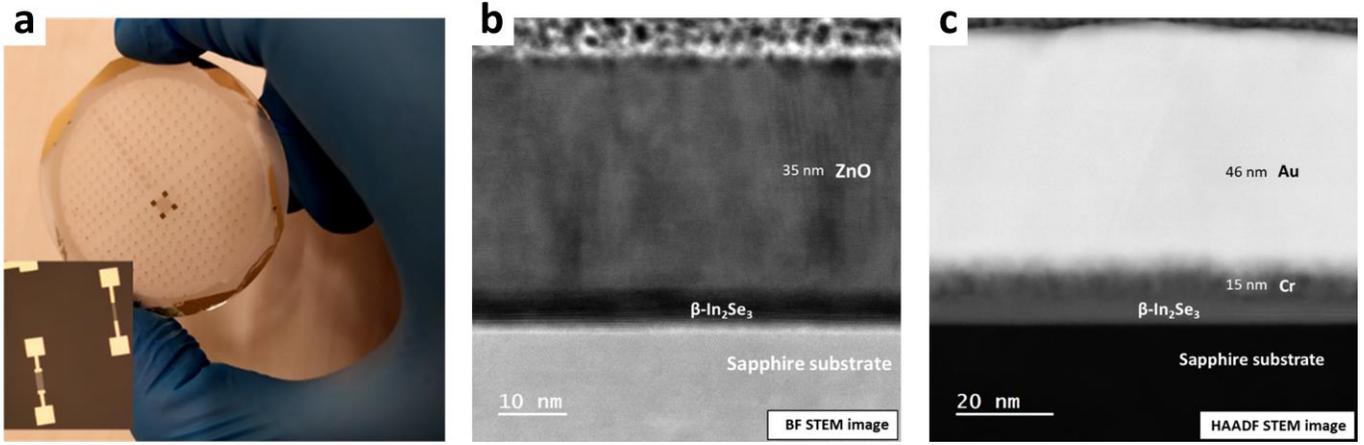

**Figure 5.** Wafer-scale device fabrication of 5 QL β-In$_2$Se$_3$. **a** Processed 2-inch *c*-sapphire wafer containing more than 250 devices. The inset shows an optical microscope image of two of the studied devices. Cross-section HAADF ac-STEM of the produced devices in **b** the channel region and **c** the contact region.

The observed dark current and photocurrent under white light for devices of both sizes are highly reproducible with only small variations across the wafer (Fig. 6a). This result demonstrates the benefit of our MBE-grown β-In$_2$Se$_3$ on 2-inch wafers, providing hundreds of identical devices. Device fabrication using exfoliated 2D materials in a one-by-one process often leads to large variation in device characteristics.[5]

The wavelength-dependent responsivity was determined using monochromatic light (Fig. 6b), from which an optical bandgap energy of $E_g$ = 1.38 eV is extracted. This value is in agreement with the theoretically calculated bandgap and that measured by other methods[36], and corresponds to the one expected for bulk β-In$_2$Se$_3$[27]. From the responsivity, we determine the external quantum efficiency (EQE) to 0.67%. This value is on the order of the EQE reported for few layers of other 2D materials, with a typical absorption of ~ 0.1% per layer.[6]

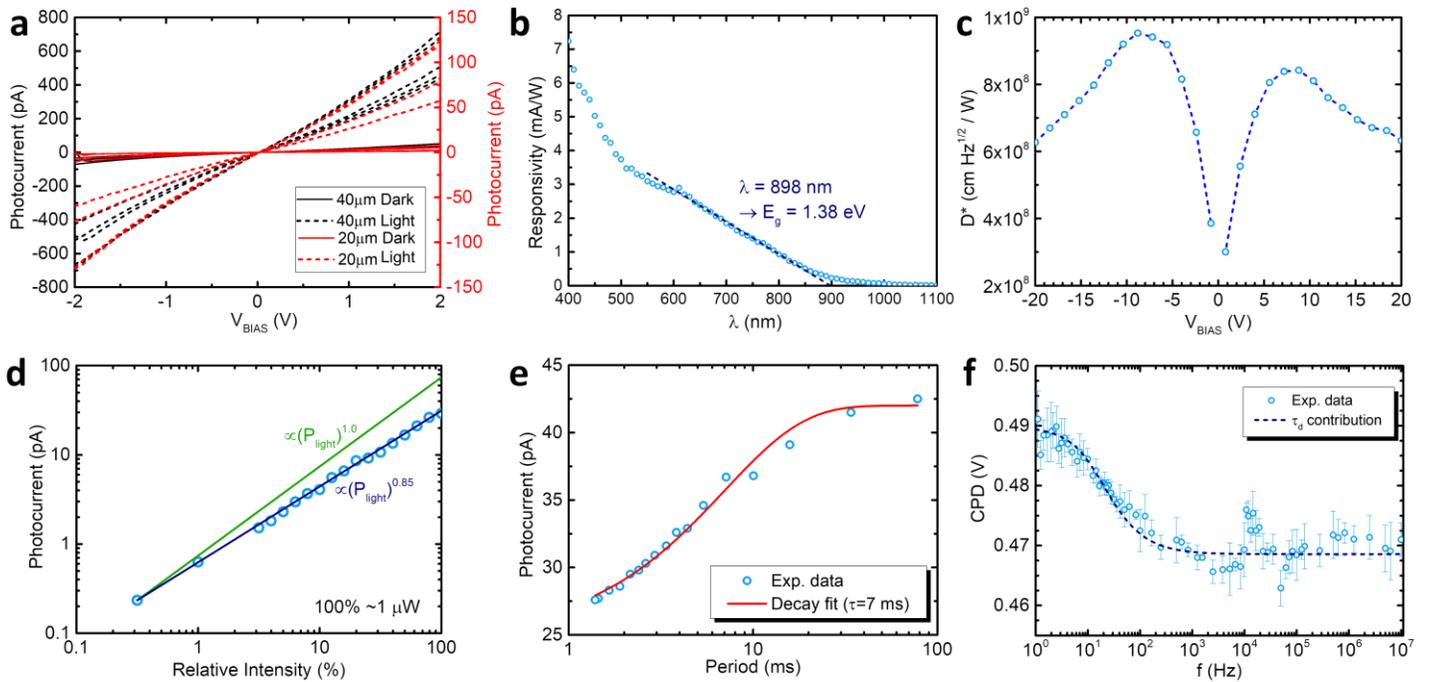

**Figure 6.** Characteristics of photodetector devices based on 5 QL β-In$_2$Se$_3$. **a** Photocurrent (dashed) and dark current (continuous) under white light for several devices with 100 μm channel length and 40 μm (black, left axis) and 20 μm (red, right axis) channel width (W). Power density of 40 W/m$^2$. **b** Spectral responsivity at $V_{bias}$ = 2 V, (W = 40 μm). Average power density of 8 W/m$^2$. **c** Specific detectivity (D*) vs. $V_{bias}$ at 550 nm illumination (W = 40 μm). **d** Photocurrent vs. light intensity for white light exhibiting a $(P_{light})^\alpha$ dependence with fit line for α = 0.85. The $(P_{light})^{1.0}$ dependence is included for reference (W = 40 μm). **e** Photocurrent response-time measurements using photocurrent intensity decay under light modulation (W = 20 μm). **f** TR-SPV results on thin (< 20 QL) β-In$_2$Se$_3$ layers grown on GaN providing an independent confirmation of the time scale of charge carrier dynamics in the β-In$_2$Se$_3$ material (without device processing).

To evaluate the carrier dynamics of the photo-generated charges in our β-In$_2$Se$_3$ material independent of the device fabrication and device properties, we performed time-resolved surface photovoltage (TR-SPV) measurements in a Kelvin probe force microscope (KPFM). The SPV, defined as the difference between the contact potential difference (CPD) under illumination and that in the dark, was measured as a function of the modulation frequency of a light source (635 nm). For these measurements, a conductive substrate is required to be able to electrically contact the β-In$_2$Se$_3$. Therefore, we developed a growth process for high quality β-In$_2$Se$_3$ on commercially-available *c*-sapphire wafers with an epitaxially grown GaN (thickness = 5 μm) film. The GaN is conductive and we were able to confirm the growth of β-In$_2$Se$_3$ with similar quality on these substrates (see Fig. S6 [Suppl. Information]), using essentially the same growth conditions. Figure 6f shows the SPV spectra as a function of the modulation frequency of the illumination. The negative SPV indicates that the material is n-type, as mostly found for indium-selenides and other 2D chalcogenides due to selenium vacancies. The major process in the SPV spectrum as a function of the modulation frequency of the light[40-42] can be well described by a single exponential decay, with a decay time of $\tau_{spv}$ = 14 ms, which is of the same order as the photoconductivity response time ($\tau_{pc}$ ~ 7 ms). A more detailed fit of the SPV spectrum (Fig. S7 [Suppl. Information]) confirms this value of $\tau_{spv}$ for the dominant process. Both time constants ($\tau_{pc}$ and $\tau_{spv}$) are in the same range of ~10 ms, indicating that the same carrier recombination process might control the response time. In the photodetector, photo-generated carriers travelling to the contacts govern the response time. In the TR-SPV, the time constant is governed by the separation of photo-excited charge carriers and their recombination. The TR-SPV results were obtained on bare β-In$_2$Se$_3$ material without any device processing and relate most likely to an intrinsic property of the thin β-In$_2$Se$_3$ layer. We therefore presume that in both methods the release of carriers from trap states in the β-In$_2$Se$_3$ material governs the time scale. A similar time scale is found in many 2D photodetectors when photoconduction is the main mechanism. Frequently, the mechanism for photocurrent generation in 2D devices is assigned to photogating, photo-thermoelectric, or others effects due to device design, and in those cases, the response time is approximately seconds.[5] Slow response times correlate with long carrier lifetimes, which also result in a large carrier diffusion length ($L = \sqrt{\tau D}$, with D the Einstein diffusion coefficient, $D = \mu k_b T/q$), leading to high gain and responsivity of the 2D device, despite its low absorptivity due to the small thickness. Therefore, there is a compromise between speed and responsivity.

The diversity of photocurrent generation mechanisms could be responsible for the wide distribution of reported responsivities[6] of this material and also between other 2D materials. The responsivity (and then D*) are found over several orders of magnitude, from few mA/W, as measured in this work, to thousands of A/W depending on the growth and device fabrication methods, and it is usual that very high responsivities are associated to slow response-times. This dependence suggests that in some of these devices, the photocurrent is generated and amplified by a slow charge-trapping at the surface, the photogating effect. This effect is expected to be stronger in cases where the device is based on exfoliated or CVD deposited flakes with small areas (<200 μm), constantly exposed to air and natural oxidation. In the devices presented in this work, the β-In$_2$Se$_3$ is exposed to the environment during a very short time and is encapsulated after processing, which guarantees that the layer is not modified prior to or during the measurements. This device processing resulted in the fastest response time reported so far for this material (see Table S1 [Suppl. Information]).[16,39,43] We also found highly reproducible results in the many devices tested. Thus, the figures of merit obtained in this work are expected to better correspond to the inherent properties of the few-layers 2D material than to uncontrollable surface or interface effects as found in single, small devices based on flakes.

In summary, we show the epitaxial growth of high-quality and single-phase β-In$_2$Se$_3$ by MBE on a 2-inch wafer scale. Thin layers down to two QLs were uniformly obtained. We fabricated photodetectors using photolithography and other standard semiconductor processing, demonstrating for the first time large-scale fabrication of β-In$_2$Se$_3$ 2D-material devices, enabling the future integration with established semiconductor technologies. Our photodetectors based on 5 QLs of β-In$_2$Se$_3$ are sensitive to wavelengths up to 898 nm (1.38 eV) and show a responsivity of 3 mA/W, peak specific detectivity (D*) of $10^9$ Jones, fast response time of ~7 ms, and external quantum efficiency (EQE) of 0.67% at 550 nm. These figures of merit are comparable to other reported 2D photodetectors for which the photocurrent generation is due to the photoconduction mechanism. We expect that our demonstration of wafer-scale deposition of the 2D material $In_2Se_3$ and its fabrication into optoelectronic devices will pave the way for an accelerated commercialization of 2D materials.

**Methods.** The $\beta$-$In_2Se_3$ growth was performed in an EVO-50 molecular beam epitaxy (MBE) system (Omicron Nanotechnology GmbH). Indium (6N) is evaporated from a Knudsen cell and selenium (5N) from a valved cracker cell. The Se is evaporated from a reservoir maintained at 285 ºC, while the flux is controlled by a valve with an adjustable aperture ranging from 0 to 8 mm. Before entering the growth chamber, larger selenium molecules are cracked by the cracker stage kept at 900 ºC. The stand-by base pressure of the MBE system is $2.6 \times 10^{-10}$ mbar and during the growth, when the Se valve is open, the pressure increases to $10^{-8}$ to $10^{-7}$ mbar. All growth processes are observed by reflection high-energy electron diffraction (RHEED, Staib Instruments), which was operated at 15 kV. Epi-ready single-side polished 2-inch c-sapphire (0001) substrates were used, with a specified and confirmed roughness of ~0.2 nm. Each substrate was annealed inside the growth chamber for 30 min at 950 ºC just prior to the growth of $In_2Se_3$.

Raman spectroscopy was measured at room temperature in a Witec alpha300 R confocal microscope, using a 50X objective lens, and a solid-sate 532 nm excitation laser. X-ray diffraction (XRD) measurements were performed in a PANalytical Xpert PRO MRD diffractometer with 5-axis cradle, standard Bragg-Brentano (BB) geometry, Cu anode X-ray tube operated at 45 kV accelerating voltage and 40 mA filament current to generate X-rays (Cu K-alpha). Soller and collimation 0.5" slits were used in the source side and detection using Soller slits and CCD detector (PiXcel) inline (1D) model. The gracing-incidence XRD (GIIXRD) (and φ-scan) was done in similar configuration (BB), but with the addition of a Goebel mirror in the source side and using open-detector (0D) for detection.

Cross-sectional sample preparation for STEM imaging was carried out by a focused ion beam (FIB) using an FEI Helios NanoLab 450S Dual Beam Focused Ion Beam with UHREM FEG-SEM. To protect the $In_2Se_3$ layers from oxidation for the STEM observation, a Mo layer was deposited by sputtering immediately after MBE growth. Additionally, a 2$^{nd}$ layer of Pt, using first the electron beam and then the ion beam, was deposited to protect the sample from Ga ion implantation and consequent damage during FIB preparation. The structural characterization was performed by STEM imaging using a FEI Titan Cubed Themis 60-300 kV microscope equipped with Probe and Image Correctors. The microscope was operated at 200 kV with a convergence angle of 21 mrad. The inner and outer collection angles used for high-angle annular dark-field (HAADF) STEM imaging were 50.5 and 200 mrad, respectively.

Atomic force microscopy (AFM) measurements were taken with a BRUKER Dimension Icon in tapping-mode using PPP-NCH (Nanosensors$^{TM}$) cantilevers with a nominal tip radius of < 20 nm, force constant of 42 N/m, and ~ 265 kHz resonance frequency. Kelvin probe force microscopy (KPFM) measurements were carried out using an ultra-high vacuum scanning probe microscope (Omicron Nanotechnology GmbH), controlled by a Nanonis controller (SPECS Zurich GmbH). Pt/Ir-coated Si cantilevers (Nanosensors$^{TM}$) were used ($f_0 \approx 168$ kHz). Amplitude modulation (AM) at the second resonance frequency of the cantilever ($f_2 \approx 1.043$ MHz) was used for the detection of the contact potential difference (CPD), where CPD = $\Phi_{sample} - \Phi_{tip}$. The work function of the tip ($\Phi_{tip}$) was calibrated using an Au reference sample. To obtain the time-resolved surface photovoltage (TR-SPV), the average SPV is measured as a function of the frequency of a modulated light source.[40-42] The surface photovoltage is defined as SPV = $CPD_{light}$ - $CPD_{dark}$. For the modulated light a fast switched diode laser (PicoQuant FSL500) was externally triggered using user-defined signal patterns, avoiding any frequencies which could lead to artifacts due to frequency mixing.[44]

Photocurrent and IV curves were measured by a Keithley 6487 picoamperemeter when in DC mode or by an AMETEK 5210 lock-in amplifier when the light was modulated by a mechanical chopper. For these experiments illumination was provided by unfiltered white halogen tungsten lamp, in combination with a colored glass filter (550 nm) or by a QuantumDesign MLS-450-300 monochromator. The light intensity was measured using a calibrated Si PIN photodiode (FDS010).

ASSOCIATED CONTENT

**Supporting Information**.
The following files are available free of charge.
Additional experimental descriptions and data (PDF)

AUTHOR INFORMATION

**Corresponding Author**
*marcel.claro@inl.int

**Author Contributions**
M.S.C. designed and executed the MBE growth and device processing, performed the Raman spectroscopy, AFM, XRD, GIIXRD, photocurrent and others device characterization. J.G. executed the HR-STEM measurements. N.N. performed the TR-KPFM. P.J.F. and S.S. supervised and coordinated the study. All the authors contributed to the data analysis, interpretation of the results, and writing the manuscript.

**Funding Sources**
This article has received support from the project Nanotechnology Based Functional Solutions (NORTE-01-0145-FEDER-000019), supported by Norte Portugal Regional Operational Programme (NORTE2020), under the PORTUGAL 2020 Partnership Agreement, through the European Regional Development Fund (ERDF). Additional support by National Funds through the Portuguese Foundation for Science and Technology (FCT) in the framework of the project "LA2D" - PTDC/FIS-NAN/3668/2014 is acknowledged. This work was supported by FCT, through IDMEC, under LAETA, project UIDB/50022/2020.

ACKNOWLEDGMENT

We would like to thank Tony Hart for the support with photocurrent measurements.

# Supporting Information

# Wafer-scale fabrication of fast two-dimensional β-In$_2$Se$_3$ photodetectors

*Marcel S. Claro [a,b]\*, Justyna Grzonka [a], Nicoleta Nicoara [a,b], Paulo J. Ferreira [a,c,d], Sascha Sadewasser [a,b]*

[a] International Iberian Nanotechnology Laboratory, Av. Mestre José Veiga, 4715-330, Braga, Portugal

[b] QuantaLab, 4715-330 Braga, Portugal

[c] Mechanical Engineering Department and IDMEC, Instituto Superior Técnico, University of Lisbon, Av. Rovisco Pais, 1049-001 Lisboa, Portugal

[d] Materials Science and Engineering Program, The University of Texas at Austin, Austin, Texas 78712, USA.

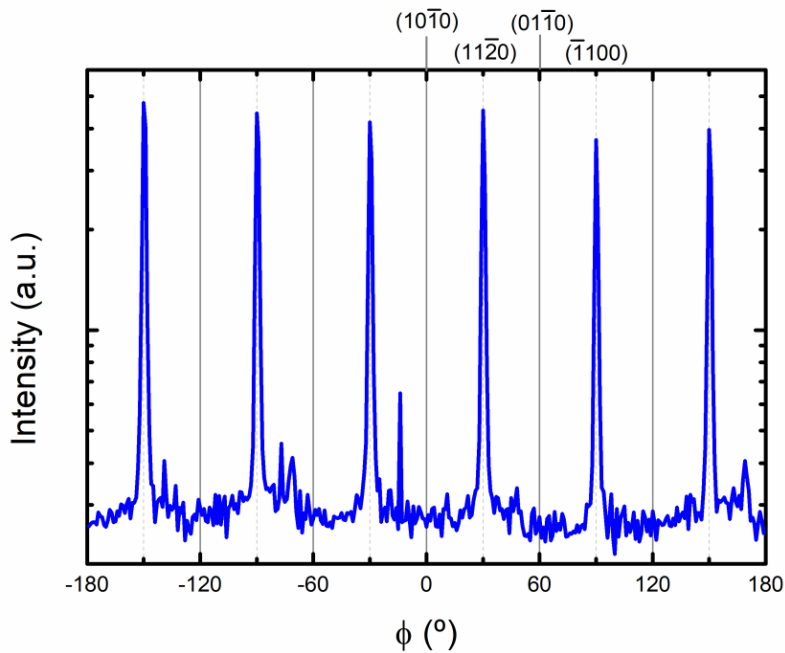

**Figure S1.** GIIXRD Φ-scan of the $(11\bar{2}0)$ peak of the 90 nm thick β-In$_2$Se$_3$ sample.

**Structure analysis of 5 QL β-In$_2$Se$_3$ on c-sapphire.**

    Thin, five QLs thick β-In$_2$Se$_3$ films were grown using a growth time of 90 s. AFM imaging indicates the same RMS roughness as that found for the thick sample (1.35 nm), but a significantly lower number of terraces is observed (Fig. S2a). These terraces are smaller in size compared to the thick sample. Around the average thickness of 5 QLs, terraces with 4 QL and some small areas with the 6$^{th}$ QL forming are observed, confirming that a continuous and uniform film of 5±1 QLs was obtained. The height of each QL is confirmed by a height line profile to be 0.95 nm (Fig. S2b). This topology, which resembles the start of layer-by-layer growth, is uniform across the 2-inch wafer and reproducible over many growth processes.

    The β-In$_2$Se$_3$ structure of the 5 QL sample is confirmed by Raman spectroscopy and XRD measurements, even though the signals are much weaker due to the thinner sample. In the XRD diffractogram (Fig. S2c), the (004) peak is the most intense and clearly visible despite the weakness and broadness typical of such thin films. The Raman spectrum (Fig. S2d) differs from the thick sample by the appearance of an amorphous-Se (a-Se) peak with exposure time (during the measurement the peak height increases every second), or when using higher laser power. The same effect was observed on similar ultra-thin chalcogenides.[1,2]

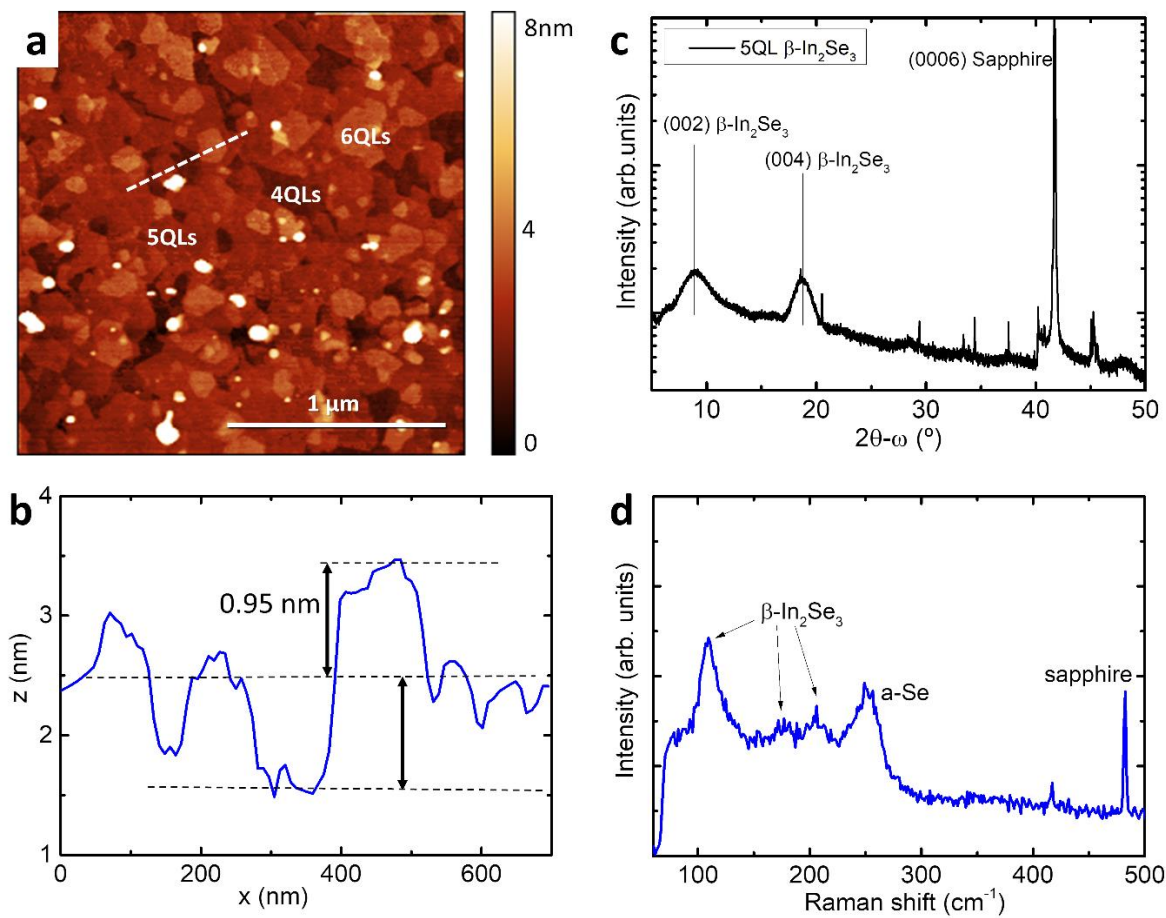

**Figure S2.** Topography and structure of a 5 QL β-In$_2$Se$_3$ sample. **a** Atomic force microscopy (AFM) of the surface showing 3 different terrace levels. 2 × 2 μm$^2$. **b** Line profile along the dashed line in a, showing terraces at three different levels, corresponding to 4, 5, and 6 QL thickness. **c** X-ray diffraction (XRD) taken immediately after growth (black). **d** Raman spectrum of 5 QL β-In$_2$Se$_3$ samples obtained at 1 mW during 4 s.

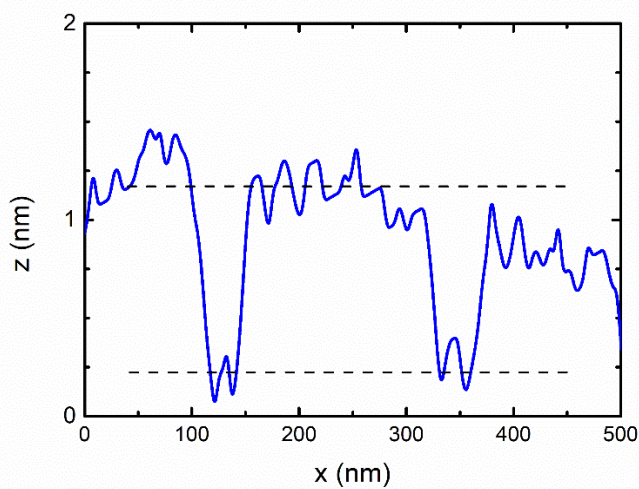

**Figure S3.** Height profile for the 2 quintuple layers (QL) sample, showing individual steps of ~0.95 nm height, corresponding to a single QL.

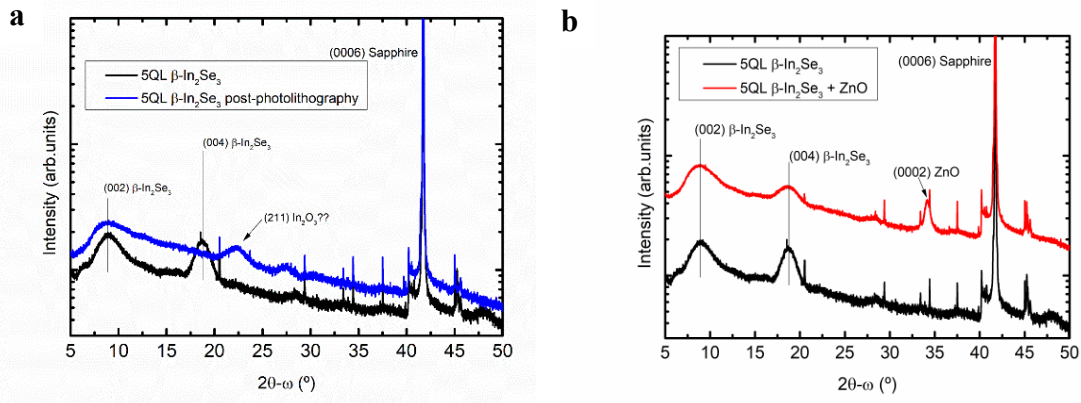

**Figure S4.** X-ray diffraction (XRD) of a 5 QLs β-In$_2$Se$_3$ samples. **a** Diffractogram taken directly after the growth (black), and after being processed by photolithography (blue) for device fabrication without passivation layer. The diffractogram shows that β-In$_2$Se$_3$ is largely converted to In$_2$O$_3$ due to the processing. **b** X-ray diffractogram taken immediately after the growth (black) and after sputter-deposition of an additional i-ZnO layer (30 nm thickness, red).

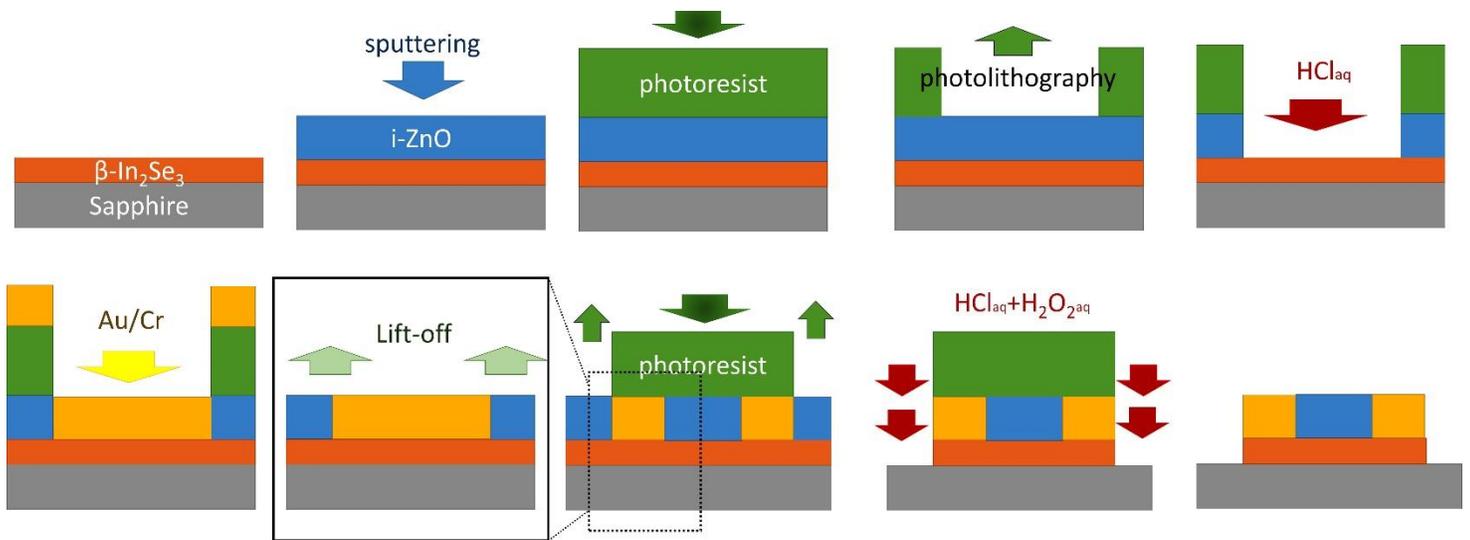

**Figure S5.** Schematics of the device fabrication sequence, starting with β-In$_2$Se$_3$ on a c-sapphire wafer and ending with photodetector devices.

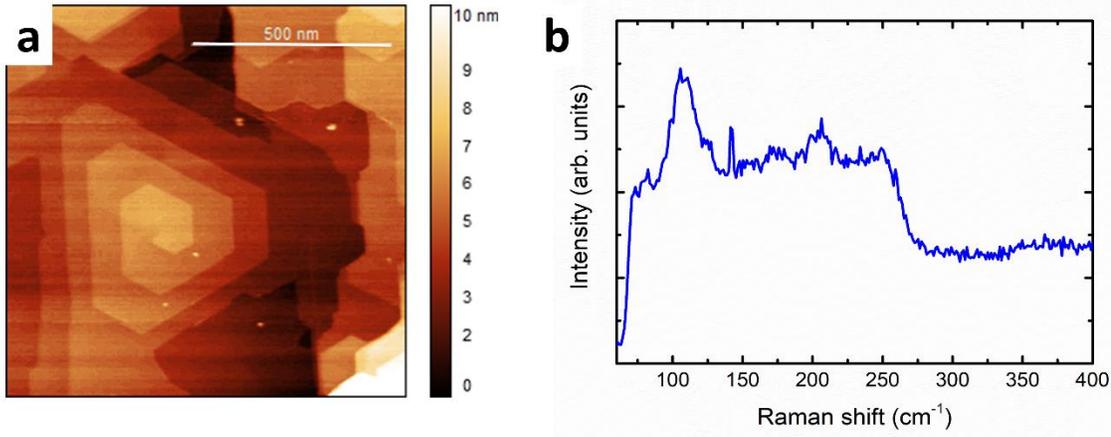

**Figure S6.** **a** Atomic force microscopy (AFM) of the surface of β-In$_2$Se$_3$ on a GaN/sapphire substrate. The nominal thickness is 20 nm. The In$_2$Se$_3$ film shows areas with β phase and others with γ phase, which can be easily distinguished in AFM and Kelvin probe force microscopy. **b** Micro-Raman spectrum showing the signature of β-In$_2$Se$_3$.

**Time-resolved surface photovoltage (TR-SPV) and photodetector response time.**

To fit the time response of the photodetector device in Fig. 6e, a simple exponential behavior was used:

$$y = y_0 + A_1 \cdot e^{-(x-x_0)/\tau} \qquad \text{(Eq. S1)}$$

The detailed analysis of the time-resolved surface photovoltage (TR-SPV) is performed following Refs. [3,4]. The time dependence of the SPV is described by three exponential terms, a build-up process with a time constant $\tau_b$, and two decay processes with time constants $\tau_{d1}$ and $\tau_{d2}$. The frequency-dependent contact potential difference (CPD) data from TR-SPV experiments with modulated light can then be described by:

$$\begin{aligned} CPD(f) = {} & CPD_{dark} + SPV_1 D \left(1 - e^{\frac{-(1-D)}{\tau_{d1} \cdot f}} e^{\frac{-D}{\tau_{b1} \cdot f}}\right) \\ & + SPV_1 (\tau_{d1} \cdot f - \tau_{b1} \cdot f)\left(1 - e^{\frac{-(1-D)}{\tau_{d1} \cdot f}}\right)\left(1 - e^{\frac{-D}{\tau_{b1} \cdot f}}\right) \\ & + SPV_2 D + SPV_2 (\tau_{d2} \cdot f)\left(1 - e^{\frac{-(1-D)}{\tau_{d2} \cdot f}}\right) \end{aligned} \qquad \text{(Eq. S2)}$$

where $CPD_{dark}$ is the in CPD measured in the dark, $SPV$ is the surface photovoltage measured under continuous illumination, $f$ is the modulation frequency, and $D$ is the illumination duty cycle.

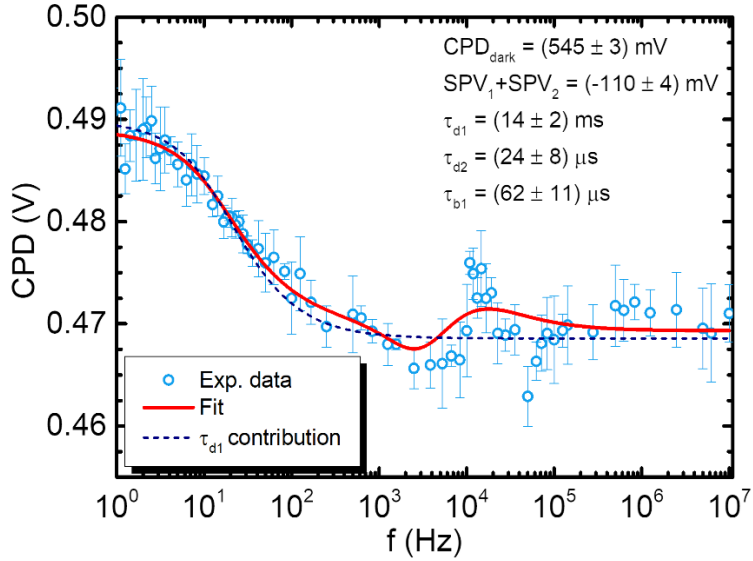

**Figure S7**. Charge carrier dynamics measured by TR-SPV on thin (< 20 QL) β-$In_2Se_3$ layers grown on GaN.

The SPV spectrum as a function of the modulation frequency of the light (Fig. S7) can be well described by Eq. (S2) with a single exponential build-up process and two exponential decay processes, with time constants $\tau_b$, $\tau_{d1}$, and $\tau_{d2}$, respectively[5-7], where the decay time constants are typically associated with the minority-carrier lifetime. A respective fit to our data is indicated by the solid line in Fig. S7 and yields the two decay time constants $\tau_{d1}$ ~ 14 ms and $\tau_{d2}$ ~ 24 µs, and a built-up time constant $\tau_{b1}$ ~ 62 µs. The dominant decay process is characterized by $\tau_{d1} = \tau_{spv}$ ~ 14 ms.

**$In_2Se_3$ photodetectors**

Table S1: Comparison of $In_2Se_3$ photodetector devices.

| Material | Preparation method | Responsivity (A/W) | Response time (ms) | Reference |
|---|---|---|---|---|
| β-$In_2Se_3$ | MBE | 0.003 | 7 | This work |
| β-$In_2Se_3$ | Exfoliation | 3.84 | 4350 | 8 |
| β-$In_2Se_3$ | Patterned CVD growth | 1650 | 1000 | 9 |
| α-$In_2Se_3$ | Exfoliation | 395 | 18 | 10 |
| α-$In_2Se_3$ | Flakes by vapor transport | 340 | 6 | 11 |
| α-$In_2Se_3$ | Exfoliation | 98000 | 9000 | 12 |
| α-$In_2Se_3$ | Transfer of CVD flakes | 0.37 | 0.2 | 13 |
| α-$In_2Se_3$ | Exfoliation | 1081 | 8 | 14 |